\documentclass[preprint,unnumsec,webpdf,modern,large,namedate]{bioinformatics}

\usepackage{graphicx} 
\usepackage{url}
\usepackage{hyperref}
\usepackage{fix-cm}
\title[Protein Structure Measurement and Visualization]{ProCaliper: functional and structural analysis, visualization, and annotation of proteins}

\author[1]{Jordan C. Rozum\ORCID{0000-0002-4356-9809}}
\author[1]{Hunter Ufford}
\author[1]{Alexandria K. Im}
\author[1]{Tong Zhang\ORCID{0000-0003-2540-2017}}
\author[2,1]{David D. Pollock\ORCID{0000-0002-7627-4214}}
\author[$\ast$,1]{Doo Nam Kim\ORCID{0000-0001-9895-7190}}
\author[$\ast$,1]{Song Feng\ORCID{0000-0003-3983-9009}}
\address[1]{\orgdiv{Biological Sciences Division}, \orgname{Pacific Northwest National Laboratory}, \orgaddress{Richland WA, 99354, USA}}
\address[2]{\orgdiv{Department of Biochemistry and Molecular Genetics}, \orgname{University of Colorado School of Medicine}, \orgaddress{Aurora, CO, USA}}
\corresp[$\ast$]{Corresponding authors: \href{song.feng@pnnl.gov}{song.feng@pnnl.gov} (SF), \href{doonam.kim@pnnl.gov}{doonam.kim@pnnl.gov} (DNK)}

\newcommand{\pkg}[1]{\texttt{#1}}

\begin{document}

\journaltitle{PREPRINT}
\DOI{PENDING}
\copyrightyear{2025}
\pubyear{2025}
\access{PREPRINT}
\appnotes{PAPER}

\firstpage{1}
\abstract{
    Understanding protein function at the molecular level requires connecting residue-level annotations with physical and structural properties. This can be cumbersome and error-prone when functional annotation, computation of physico-chemical properties, and structure visualization are separated.
    To address this, we introduce \pkg{ProCaliper}, an open-source Python library for computing and visualizing physico-chemical properties of proteins. It can retrieve annotation and structure data from UniProt and AlphaFold databases, compute residue-level properties such as charge, solvent accessibility, and protonation state, and interactively visualize the results of these computations along with user-supplied residue-level data. Additionally, \pkg{ProCaliper} incorporates functional and structural information to construct and optionally sparsify networks that encode the distance between residues and/or annotated functional sites or regions.
    The package \pkg{ProCaliper} and its source code, along with the code used to generate the figures in this manuscript, are freely available at \url{https://github.com/PNNL-Predictive-Phenomics/ProCaliper}.
}

\maketitle

\section{Introduction}
Protein data is abundant and readily accessible in several large databases such as UniProt and AlphaFold~\citep{uniprotconsortiumUniProtUniversalProtein2018,jumperHighlyAccurateProtein2021,varadiAlphaFoldProteinStructure2022}. These contain computed or measured structures, functional annotations, and metadata describing post-translational modifications (PTMs)~\citep{debrevernCurrentStatusPTMs2022}. At the same time, high-throughput experiments can now probe residue-level responses to systemic perturbations~\citep{gluth2024integrative, yu_mass_nodate, dorig_global_2024}.Understanding the molecular mechanisms underpinning changes in protein function often requires coupling these data with computation or inference of protein physico-chemical properties~\citep{philippMutationExplorerWebserverMutation2024,bludauStructuralContextPosttranslational2022,cappelletti_dynamic_2021,medvedevLeveragingAIExplore2025,kimArtificialIntelligenceTransforming2025}. 
However, integrating multiple databases and novel experimental results remains cumbersome and error-prone in practice.  This can require multiple steps of interconversion between different structures and metadata formats. While there has been recent innovation in protein-ligand interaction mapping tools~\citep{pandaPritampanda15PandaMap2025}, user-friendly, integrated, and cross-platform tools for visualizing property-mapped protein structures and molecular networks are needed to make protein structural modeling accessible to all.

To address these challenges, we developed the open-source Python library \pkg{ProCaliper}.
It has integrated support for interfacing with UniProt, either through pre-downloaded tables or through the UniProt API via the \pkg{UniProtMapper} tool~\citep{DavidAraripeUniProtMapperPython}. It can store and process protein structures in PDB format. These structure files can be automatically fetched from the AlphaFold database, a user-specified alternate database, or a local file. We implemented several methods to calculate physico-chemical properties at the residue level, such as charge, solvent accessible surface area (SASA), acid dissociation constants (typically denoted $pK_a$), and more. In addition, \pkg{ProCaliper} can construct contact maps and region-to-region distance networks to describe the global structure of a protein. To sparsify these networks, \pkg{ProCaliper} includes functions for thresholding sparsification and distance backbone methods~\citep{simasDistanceBackboneComplex2021}. We support importing custom residue-level and protein-level data, which straightforwardly integrates experimental measurements into a cohesive Python representation of individual proteins. We provide methods for exporting these objects to allow for simple conversion to tabular formats that popular Python data frame packages can read or that can be written to disk (e.g., as Comma Separated Values-CSV files). We also provide tools for interactive visualization in IPython notebooks.
In this application note, we describe these features in detail and conduct an example analysis of the human heat shock protein HSP90$\alpha$ to showcase their utility.

\section{Features}
See \url{https://github.com/PNNL-Predictive-Phenomics/ProCaliper} for example usage and output. Here we describe available features.

\emph{Data Import.}
\pkg{ProCaliper} can import protein data and interface with the UniProt databases. At \url{uniprot.org}, UniProt provides tables in a standardized tab-separated value (TSV) format that \pkg{ProCaliper} natively recognizes. Users also have the option to directly download protein metadata from UniProt using the UniProtMapper API within \pkg{ProCaliper} by specifying a list of UniProt identifiers. UniProt provides protein-level annotations that indicate residue-level information such as binding sites, active sites, known ligands, and secondary structure annotations. During import, \pkg{ProCaliper} automatically parses these annotations and assigns attributes to individual residues where appropriate. At the residue level, \pkg{ProCaliper} stores data in a custom residue annotation object that ensures consistency between fetched data sets and user-supplied data (e.g., from novel PTM measurements).

\emph{Structure Calculation and Extraction.}
Several online repositories provide protein structure in the form of PDB files. By default, \pkg{ProCaliper} fetches PDB files from the AlphaFold database, but the user may specify an alternative online source for structure information or provide a local PDB file instead. We provide methods that compute residue- and protein- level features from protein structure and store them within the Protein object. Currently available methods are summarized in Table~\ref{tab:structure_methods}. Provided PDB files need not be complete; \pkg{ProCaliper} has support for reading in measured structures that do not capture the entire protein, or which may contain heteroatoms.
\renewcommand{\arraystretch}{1.0} 
\begin{table*}[]
    \caption{Structure-derived features supported using built-in \pkg{ProCaliper} methods. Additional features can be computed externally and are easily imported.}
    \centering
    \tiny
    \begin{tabular}{p{3cm}|p{4cm}|p{4cm}|p{5cm}}
         Physico-chemical property & Method(s) & Implementation Reference & Notes 
         \\\hline
         Active sites & Experiment data extraction
         & \cite{uniprotconsortiumUniProtUniversalProtein2018} & From UniProt
         \\
         Binding properties & Experiment data extraction 
         & \cite{uniprotconsortiumUniProtUniversalProtein2018} & From UniProt
         \\
         PTM sites & Experiment data extraction 
         & \cite{uniprotconsortiumUniProtUniversalProtein2018} & From UniProt
         \\
         Region annotations & Experiment data extraction
         & \cite{uniprotconsortiumUniProtUniversalProtein2018} & From UniProt
         \\
         Custom residue annotations & Not Applicable
         & NA & From user input
         \\
         Charge & gasteiger, mmff94, eem, qeq, qtpie
         & \cite{oboyleOpenBabelOpen2011} & Uses \pkg{obabel}\\
         predicted Local Distance Difference Test (pLDDT) & Prediction confidence metric extracted from AlphaFold predicted PDB file& \cite{varadiAlphaFoldProteinStructure2022} & Notated at  b-factor column in PDBs from experiment.\\
         Acid dissociation ($pK_a$) & PROPKA, pKAI, pypKa & \cite{olssonPROPKA3ConsistentTreatment2011,sondergaardImprovedTreatmentLigands2011,reis2020jcim,pkai} & As pypKa has proprietary dependencies, only PROPKA is installed by default
         \\
         Protonation state & Method from pypKa 
         & \cite{reis2020jcim} & Computed from $pK_a$
         \\
         SASA & ShrakeRupley & \cite{cockBiopythonFreelyAvailable2009} & Uses \pkg{biopython}; available at atomic-level
         \\
         Sulfur distance & Direct calculation in Python & Not Applicable & Used to identify disulfide bonds\\
         Secondary structure & Not Applicable
         & \cite{uniprotconsortiumUniProtUniversalProtein2018} & From UniProt
         \\
         Distance and proximity networks & Euclidean distance calculation & Not Applicable & Includes contact map as a special case. Support for region-to-region distance networks.
         \\
         Sparsified distance networks & Thresholding, Euclidean metric backbone & \cite{simasDistanceBackboneComplex2021} & Other backbones supported for advanced users.
         \\
         \hline
         
    \end{tabular}
    \label{tab:structure_methods}
\end{table*}

\emph{Data Export and Visualization.}
Protein objects in \pkg{ProCaliper} can be compressed and stored in binary format using Python's built-in \pkg{pickle} module for fast file reading and writing. We also provide methods to export \pkg{ProCaliper} objects in a tabular format that is human readable and can be directly imported into popular data frame libraries such as \pkg{pandas} and \pkg{polars} for further statistical analysis and visualization. The \pkg{ProCaliper} objects can be written to disk in plain text or as a spreadsheet file. Protein structures can be exported to \pkg{biopython}
or \pkg{biopandas}
formats. We also support visualization using IPython widgets via the \pkg{nglview} library~\citep{nguyenNGLviewInteractiveMolecular2018}; \pkg{ProCaliper} has methods that embed the protein structure in a 3-D interactive view in Jupyter notebooks or Visual Studio Code. It also provides methods to facilitate visualizing residue-level information within these interactive views by automatically converting data to \pkg{nglview}-compatible color schemes.

\section{Application: Heat Shock Protein HSP90$\alpha$} 

\begin{figure*}
    \centering
    \begin{tabular}{cc}
    & \multirow{3}{*}{g)\includegraphics[width=0.38\linewidth]{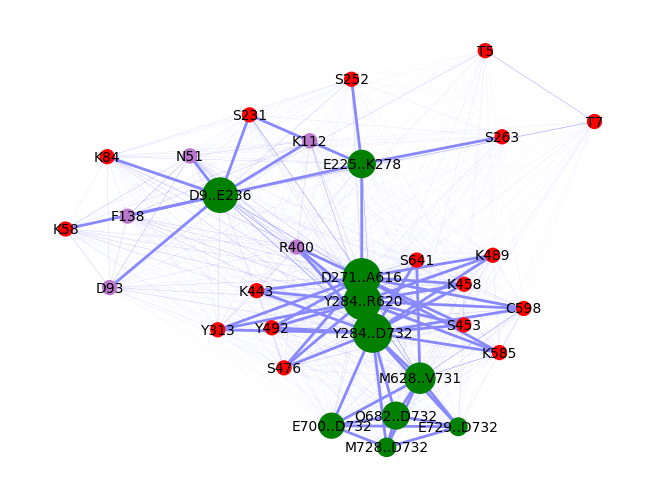}}\\
    a)\includegraphics[width=.18\linewidth]{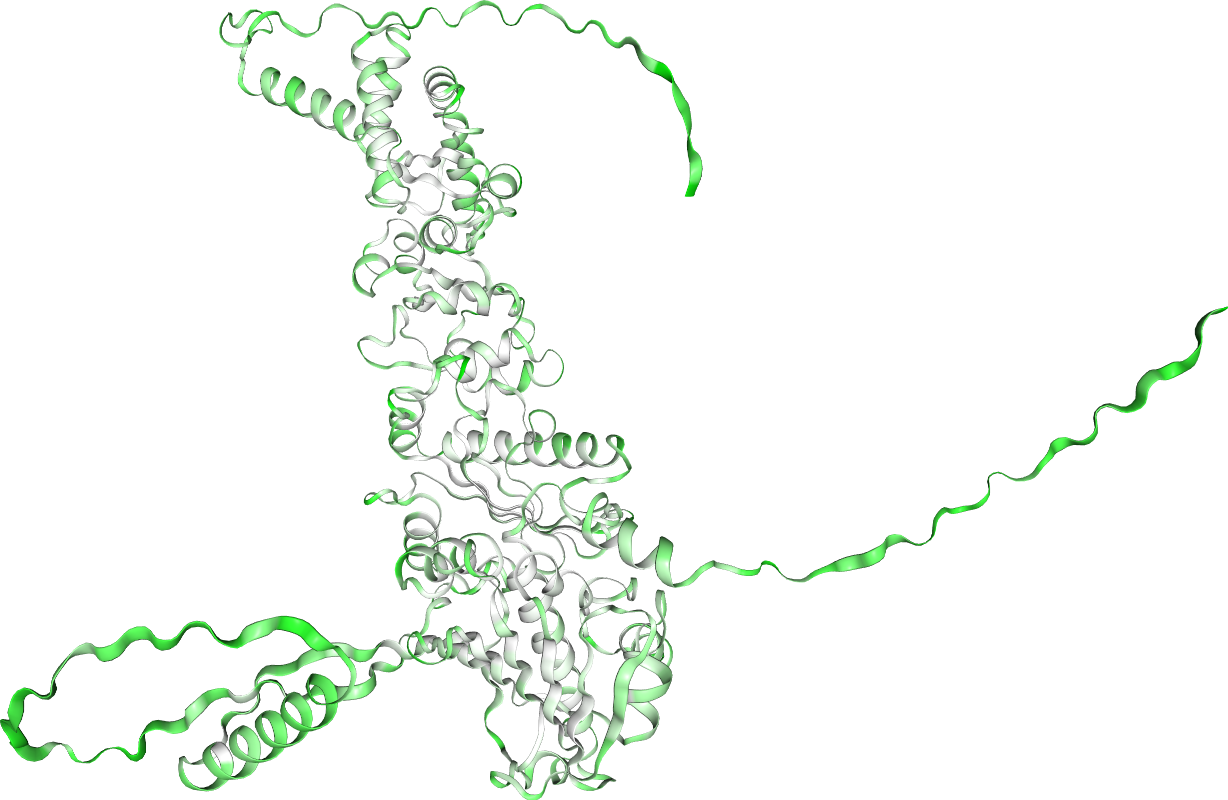}
    b)\includegraphics[width=.18\linewidth]{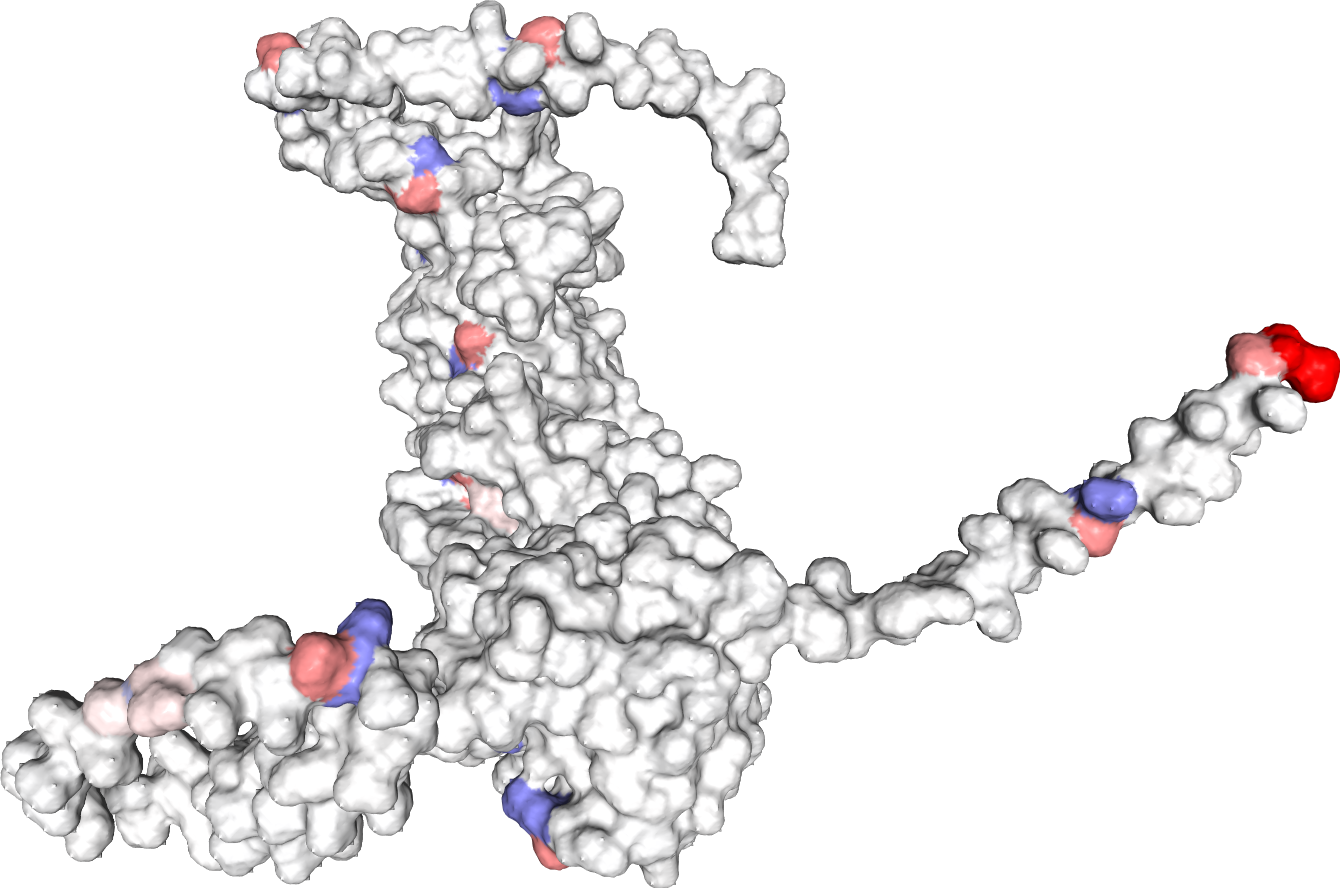}
    c)\includegraphics[width=.18\linewidth]{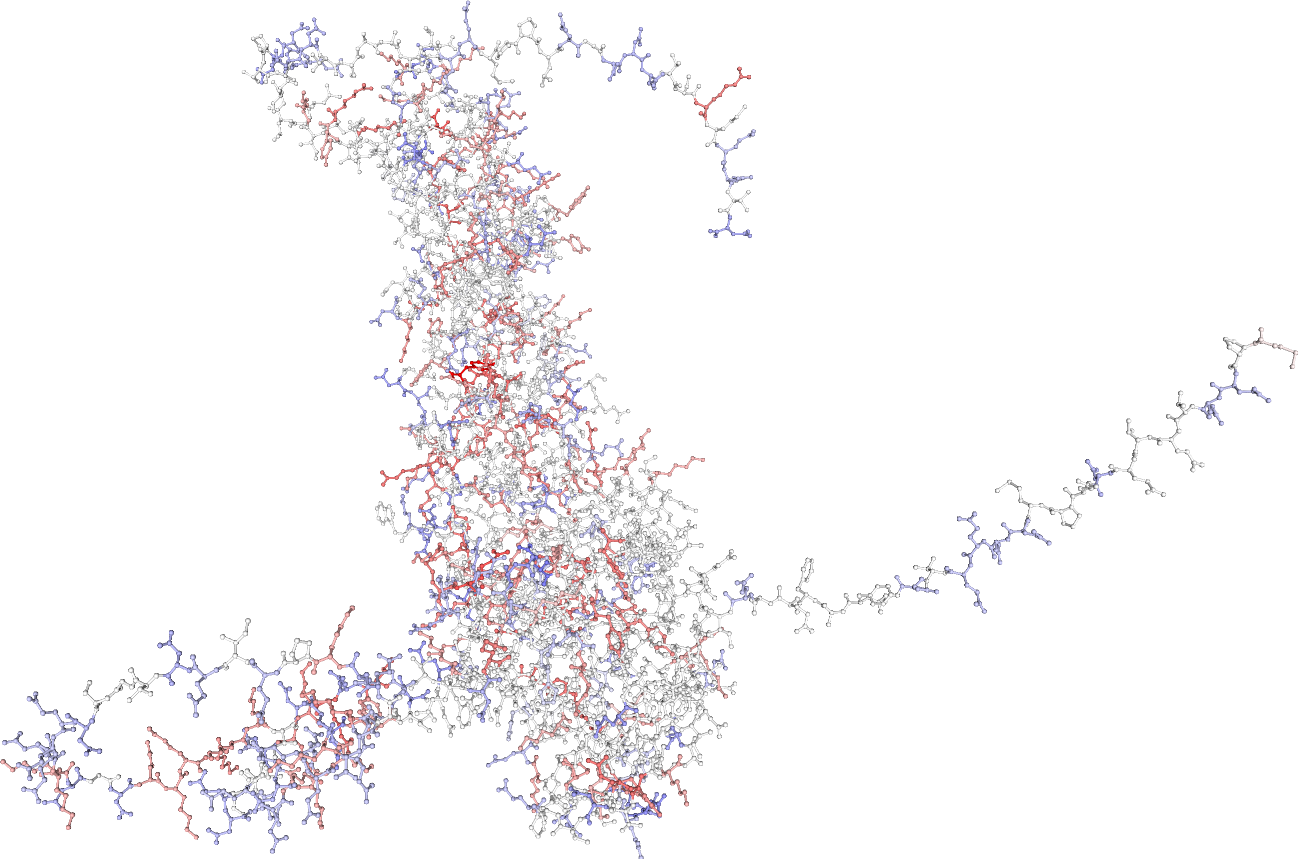}
    &\\
    d)\reflectbox{\rotatebox[origin=c]{120}{\includegraphics[width=0.18\linewidth]{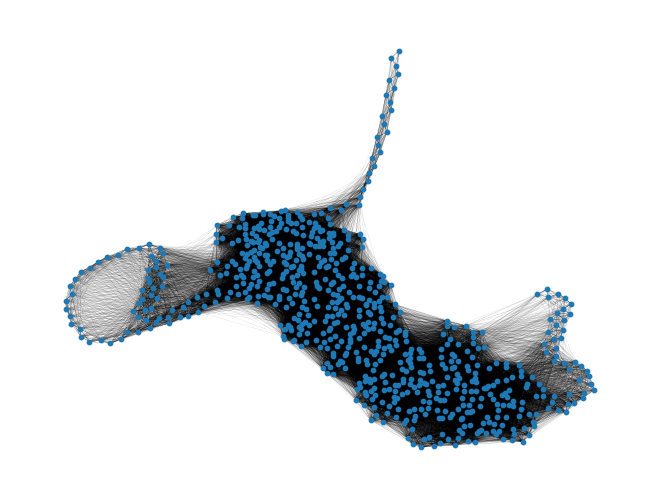}}}
    e)\reflectbox{\rotatebox[origin=c]{180}{\includegraphics[width=0.18\linewidth]{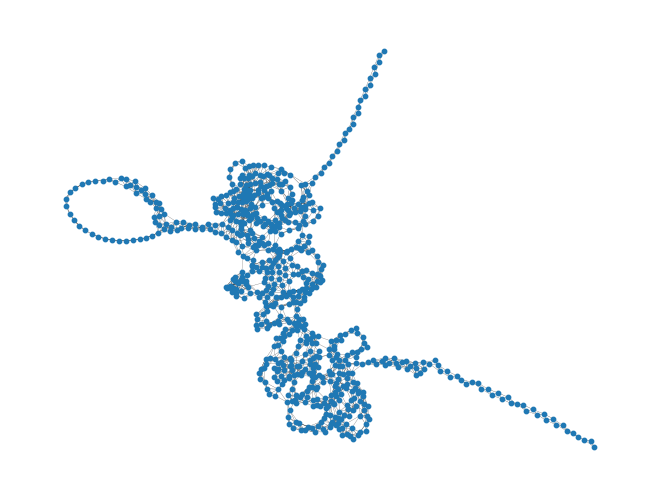}}}
    f)\reflectbox{\rotatebox[origin=c]{180}{\includegraphics[width=0.18\linewidth]{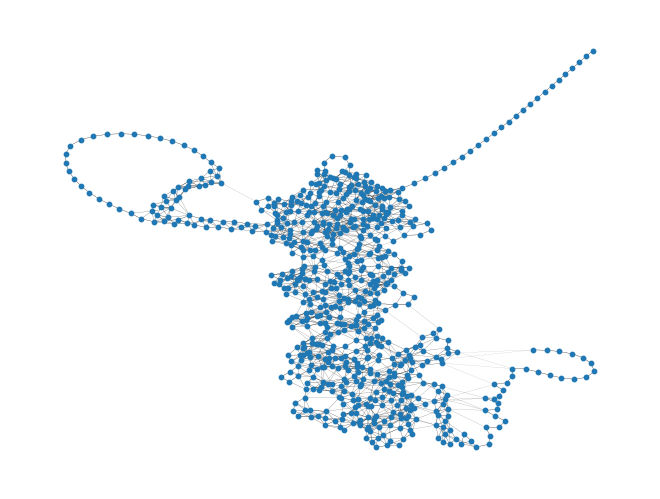}}}
    &\\
    \end{tabular}
    \caption{Visualizations of HSP90$\alpha$ using \pkg{ProCaliper}. \textbf{a-c} Interactive structure visualizations. \textbf{a)} Cartoon diagram with green intensity representing higher SASA. \textbf{b)} Surface representation colored by charge (red and blue for positive and negative, respectively). \textbf{c)} Ball and stick representation where residues are colored by $pK_a$ values (red for higher than 7, blue for lower than 7). \textbf{d-f} Distance networks plotted using the Kamada-Kawaii layout algorithm, edge thickness is inversely proportional to distance.\textbf{d)} Distance network (69,274 edges) built using a distance cutoff of 30 \AA. In \textbf{e)}, a much lower threshold (7 \AA) is used to sparsify the distance network, resulting in a contact map with 2,633 edges. In \textbf{f)}, the Euclidean backbone is used to sparsify the 30 \AA~ cutoff distance network, resulting in a network with 2,275 edges. The Euclidean backbone preserves more long-range connections and global structure than the slightly denser 7 \AA~ threshold network. \textbf{g)} Region distance graph. Annotated regions (protein interaction regions and motifs) are shown in green, ATP binding sites are shown in purple, and PTM sites are shown in red. Nodes are labeled by residue(s) and node size is proportional to the cube root of the number of residues represented by each node. Edge thickness is inversely proportional to distance plus 1 \AA, to avoid divergence. Network layout is computed using the Fruchterman-Reingold algorithm with $k=10$ and 1,000 iterations.}
    \label{fig:super-viz}
\end{figure*}

We used the human heat shock protein HSP90$\alpha$ to showcase the functionality of \pkg{ProCaliper}. HSP90$\alpha$ is a chaperone protein involved in heat stress response, protein degradation, cell cycle control, hormone signaling, and apoptosis~\citep{stetzDissectingStructureEncodedDeterminants2018,hoterHSP90FamilyStructure2018}. It is implicated in cancer metastasis and neurodegenerative diseases, possibly due to its role in protecting proteins that can degrade the extracellular matrix~\citep{yangRoleAcetylationExtracellular2008}. It has three primary domains: an N-terminal ATPase domain, an ATPase-activated middle domain that binds co-chaperones and client proteins, and a C-terminal dimer-formation domain.
We downloaded the UniProt annotations for HSP90$\alpha$ (UniProt ID P07900) and the AlphaFold-predicted structure using \pkg{ProCaliper}. Retreived annotations include binding sites, active sites, region annotations, and curated PTM sites. We then used \pkg{ProCaliper} to compute SASA, charge distribution, and $pK_a$ values for each atom and amino acid residue. We visualize these quantities in three different structure representations in an interactive IPython environment with \pkg{ProCaliper} using its built-in support for \pkg{nglview} (Figure~\ref{fig:super-viz}, panels a-c).

In panels d-f of Figure~\ref{fig:super-viz}, we show the analysis of the residue distance network with \pkg{ProCaliper} in HSP90$\alpha$. The distance network function constructs \pkg{networkx} weighted undirected graph
whose nodes correspond to the residues of an input protein and whose edge weights represent the distance (in \AA) between residues in the protein's structure. A configurable cut-off value for distances can be specified. For example, contact maps are typically constructed using a cutoff between 6 \AA~ and 12 \AA. Removing edges in a distance network by lowering this threshold can hide longer-range relationships between residues that may be important, e.g., for forming a binding pocket. In such cases, alternate methods for sparsifying a distance network are desirable, so we have implemented the Euclidean backbone method of~\cite{simasDistanceBackboneComplex2021} in \pkg{ProCaliper}. This removes any edge with weight $d$ if there is a path between its endpoints with edge weights $d_i$ and $\sqrt{\sum_i(d_i)^2}<d$. Distance backbone sparsification methods such as this cannot disconnect portions of the network~\citep{simasDistanceBackboneComplex2021} and have been used to focus analysis on salient connections in mathematical, social, and biological 
settings~\citep{rozumUltrametricBackboneUnion2024,correiaContactNetworksHave2023}.
In HSP90$\alpha$, the Euclidean backbone reduces the distance network (with initial cutoff of 30 \AA) from 69,274 edges to 2,275 edges. It preserves more long-range relationships and global structure than a thresholding reduction of 7 \AA, which removes a similar number of edges.


In \pkg{ProCaliper}, we extend the distance network concept to annotated regions and regulatory sites. In the example of HSP90$\alpha$, we highlight protein binding regions and motifs, ATP binding sites, and PTM sites (Figure~\ref{fig:super-viz}, panel g) as annotated in UniProt, but \pkg{ProCaliper} is flexible enough to incorporate user-provided annotations as well. The NR3C1-interaction region 9-236 in the N-terminal domain contains various PTM sites and ATP binding sites, and partially overlaps the disordered region 225-278. Middle domain regions (284-620, 271-616, and the interdomain region 284-732) contain various PTMs thought to play a role in chaperone binding~\citep{stetzDissectingStructureEncodedDeterminants2018}. Various regions in the C-terminal domain (those containing residues 628 through 732) overlap with one another to form a clique. There is only one nearby PTM, phosphorylation of S641. To the best of our knowledge, the functional role of this PTM, if such a function exists, remains unknown~\citep{backePosttranslationalModificationsHsp902020}.


\section{Discussion}

We have presented \pkg{ProCaliper}, a cross-platform, open-source Python library designed to facilitate interaction with the UniProt and AlphaFold databases, flexibly allow users to incorporate custom annotations and residue-level data, and to compute residue- and protein-level structural properties. It includes methods for computing residue and atom charges, acid dissociation constants, and SASA. It also provides tools for constructing and analyzing region-region distance networks (and contact maps as a special case), which represent protein structure in matrix form and have machine-learning applications~\citep{zhengDeeplearningContactmapGuided2019}. We have highlighted several of these features in the human HSP90$\alpha$ protein. We used \pkg{ProCaliper} to automatically download annotation and structure data, compute and visualize physical properties, and construct various distance networks.
Aside from its utility in studying an individual protein, \pkg{ProCaliper} also aims to facilitate interpretation of PTM proteomics data by providing a unified framework for fetching, storing, processing, and visualizing protein data at the single-residue level. Our future work will leverage \pkg{ProCaliper} to better understand the molecular mechanisms involved in protein signaling.

\section{Acknowledgments}
The research described in this paper is part of the Predictive Phenomics Initiative at Pacific Northwest National Laboratory and conducted under the Laboratory Directed Research and Development Program. Pacific Northwest National Laboratory is a multiprogram national laboratory operated by Battelle for the U.S. Department of Energy under Contract No. DE-AC05-76RL01830.

\bibliography{references}
\end{document}